# The Zeeman Effect in Finance

## *Libor Spectroscopy and Basis Risk Management*

**Marco Bianchetti**[1]
Market Risk Management, Intesa Sanpaolo,
Piazza Paolo Ferrari 10, 20121 Milan, Italy,
e-mail: marco.bianchetti[AT]intesasanpaolo.com.

## Abstract

Once upon a time there was a classical financial world in which all the Libors were equal. Standard textbooks taught that simple relations held, such that, for example, a 6 months Libor Deposit was replicable with a 3 months Libor Deposits plus a 3x6 months Forward Rate Agreement (FRA), and that Libor was a good proxy of the risk free rate required as basic building block of no-arbitrage pricing theory.

Nowadays, in the modern financial world after the credit crunch, some Libors are more equal than others, depending on their rate tenor, and classical formulas are history. Banks are not anymore “too big to fail”, Libors are fixed by panels of risky banks, and they are risky rates themselves.

These simple empirical facts carry very important consequences in derivative’s trading and risk management, such as, for example, basis risk, collateralization and regulatory pressure in favour of Central Counterparties. Something that should be carefully considered by anyone managing even a single plain vanilla Swap.

In this qualitative note we review the problem trying to shed some light on this modern animal farm, recurring to an analogy with quantum physics, the Zeeman effect.

---

[1] The author gratefully acknowledges stimulating discussions and comments by L. Cefis, M. Morini, M. Trapletti and colleagues at Market Risk Management and Fixed Income trading desks. The views expressed here are those of the author and do not represent the opinions of his employer. They are not responsible for any use that may be made of these contents in any circumstance.

# Table of contents

## 1. The Zeeman effect in physics

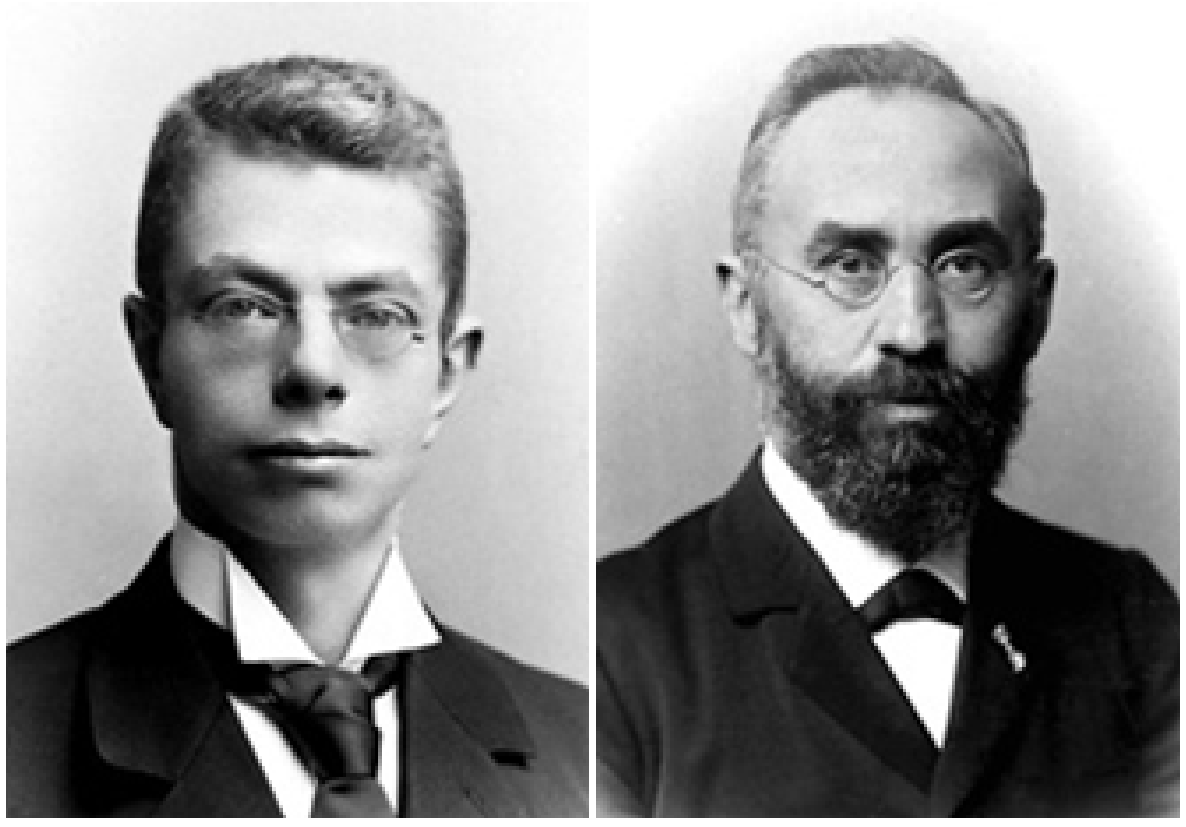

**Figure 1:** Pieter Zeeman (left) and Hendrik A. Lorentz (right).

Once upon a time in Leiden, in summer 1896, an unknown 31 years old Dutch physicist, Pieter Zeeman (see Figure 1 and ref. [1]), disobeyed his director and used the university laboratory equipment to study the effect of magnetic fields on light sources, such as common salt in a flame. He was punished for his indiscipline, but just six years later he was rewarded, jointly with Hendrik Lorentz, with the 1902 Nobel Prize in Physics for the discovery of what is universally known as the *Zeeman effect*.

**What is it ?** Atoms and molecules absorb and emit light at discrete frequencies, called *spectral lines,* which are unique and distinctive characteristic of the chemical composition of that atom or molecule. Common salt, for instance, contains Sodium which is characterized by two distinct yellow spectral lines, conventionally called $D_1$ and $D_2$, at 589.00 nm and 589.59 nm, respectively (see Figure 2, upper half [2]), that anyone can observe in sodium lamps, or just throwing a pitch of salt on a flame. If an atom or a molecule is exposed to a static magnetic field, its spectral lines are observed to split into groups, called *Zeeman multiplets* (Figure 2, lower half). Spectroscopic data of atoms and molecules were among the most comprehensive and accurate experimental observations available in Zeeman's times, thanks to the relative simplicity of the required technology.

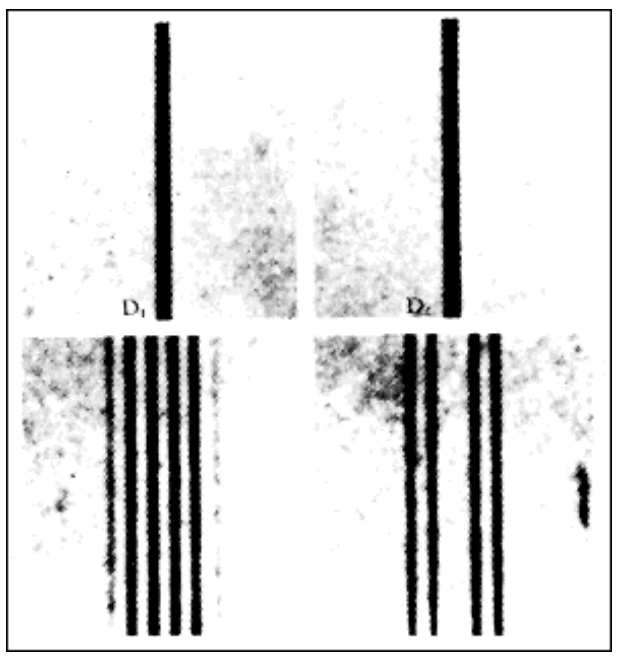

**Figure 2:** original Zeeman's picture of the Zeeman effect in Sodium.

**Why?** A few years after the discovery of Zeeman's effect, physicists discovered that atomic and molecular electrons occupy quantized states, each characterized by a unique distinctive set of quantum numbers and by a discrete energy. Spectral lines of light emission and absorption correspond to electronic transitions between different discrete energy states, thus occurring at discrete wavelengths. When such a quantum system displays certain mathematical symmetry properties (e.g. rotational invariance), some distinct electronic states with different quantum numbers may have the same energy, a situation called *quantum degeneracy*. Hence, the corresponding spectral lines have the same wavelength, and are experimentally observed as single lines. The external magnetic field, imposing a preferred direction, breaks the symmetry of the electronic states, because of the different interaction with atomic electrons characterised by different quantum numbers, thus breaking the quantum degeneracy and splitting the states' energy up and down. The corresponding spectral lines split as well, and are experimentally observed as multiple lines. See Appendix A and ref. [2] for more details.

## 2. The Zeeman effect in finance

Both Zeeman and Lorentz would be, probably, quite surprised to know that in another summer, 111 years later, a similar effect has been observed in finance. In fact, at the beginning of the credit crunch crisis in August 2007, significant splits suddenly appeared between Libor rates with different tenors quoted on the market.

**What is it ?** In Figure 1 we show the historical series of the market fixings of two interest rates, Euribor and Eonia, over a three-months (3M) period. The Euribor 3M quote is the interest rate associated with a Deposit starting at spot date (today + 2 business days) and maturing three months later (see eq. (29) and Figure 5 in Appendix B). The Eonia 3M quote is the swap rate of an Overnight Indexed Swap (OIS) with the same start and maturity dates, a fixed leg versus an Eonia indexed leg daily compounded over three months, both with a single final payment (see eq. (31) and Figure 6 in Appendix B). The difference between these two rates is called Euribor-Eonia 3M basis. We observe in Figure 1 that the two rates were essentially coincident until August 2007, when they suddenly diverged, reaching a peak of 200 basis points at the Lehman default on September 2008. This is the reason why August 2007 is conventionally indicated as the beginning of the credit crunch crisis, at least in the interest rate market. The rate correlation is usually very highly positive, but it breaks down, even below zero, in the most turbulent periods.

---

[2] 1 nm (nanometer) = $10^{-9}$ m (meters).

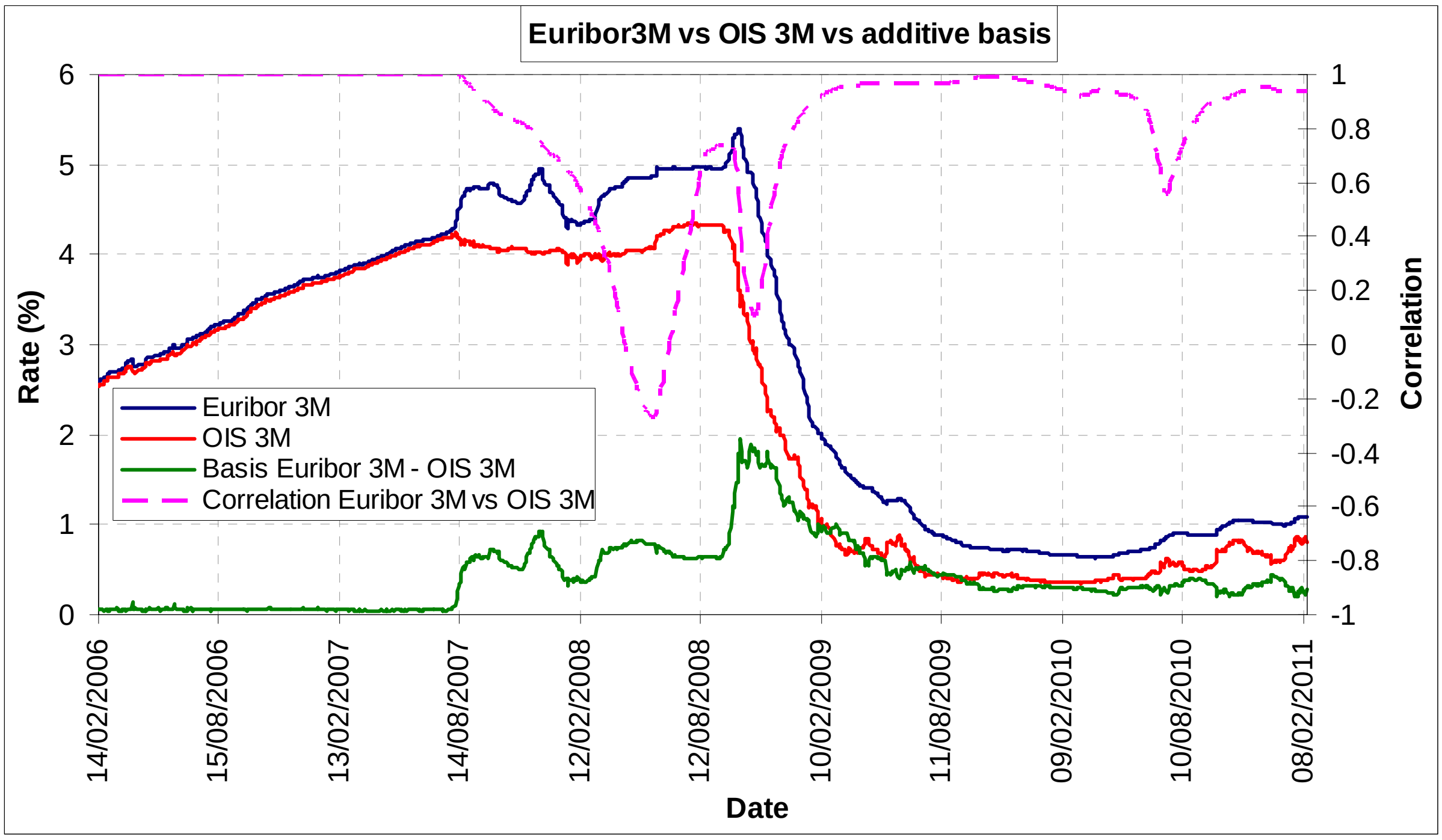

**Figure 1:** divergence between Euribor 3 months and OIS 3 months rates, and their correlation (1 year window, right scale).

Another example is reported in Figure 2, where we compare the term structures (from 1Y to 30Y maturity) of the Basis Swaps with all the possible combinations of the most common Euribor tenors (1d or over night, 1M, 3M, 6M, 12M) quoted on the market as of 30 June 2011 (see eq. (32) and Figure 7 in Appendix B).

Overall, a common pattern is observable: the larger the tenor difference, the larger the basis spread. For instance, looking at the Eonia series in Figure 2 (solid lines), we observe that the basis spread, for each maturity, increases with increasing Euribor tenor. Also these basis spreads were close to zero before the credit crunch in August 2007. Similar patterns are observed for other Libor derivatives and currencies. We have called this effect "*market segmentation*" [16]. Pushing the analogy with physics, we may refer to these empirical observations as "*Libor spectroscopy*".

**Why ?** Libor, Euribor and Eonia rates are associated with *Deposit contracts*, unsecured loans with different maturities (called tenors) used by primary banks to borrow funds on the interbank money market in different currencies (see Appendix B). Before the credit crunch, primary banks were considered, for a variety of reasons, to carry a negligible default and liquidity risk, as reflected, for example, in their small Credit Default Swap (CDS) spreads[3], able to raise infinite liquidity on the market. This is called the classical "too big to fail" paradigm.

[3] Credit Default Swaps (CDS) are Swaps constituted by a protection leg paying the receiver upon default of a given reference entity (e.g. a corporate, a bank or a sovereign), versus a floating Libor leg plus a spread (called CDS spread). The larger the default probability of the reference entity, the larger the equilibrium CDS spread.

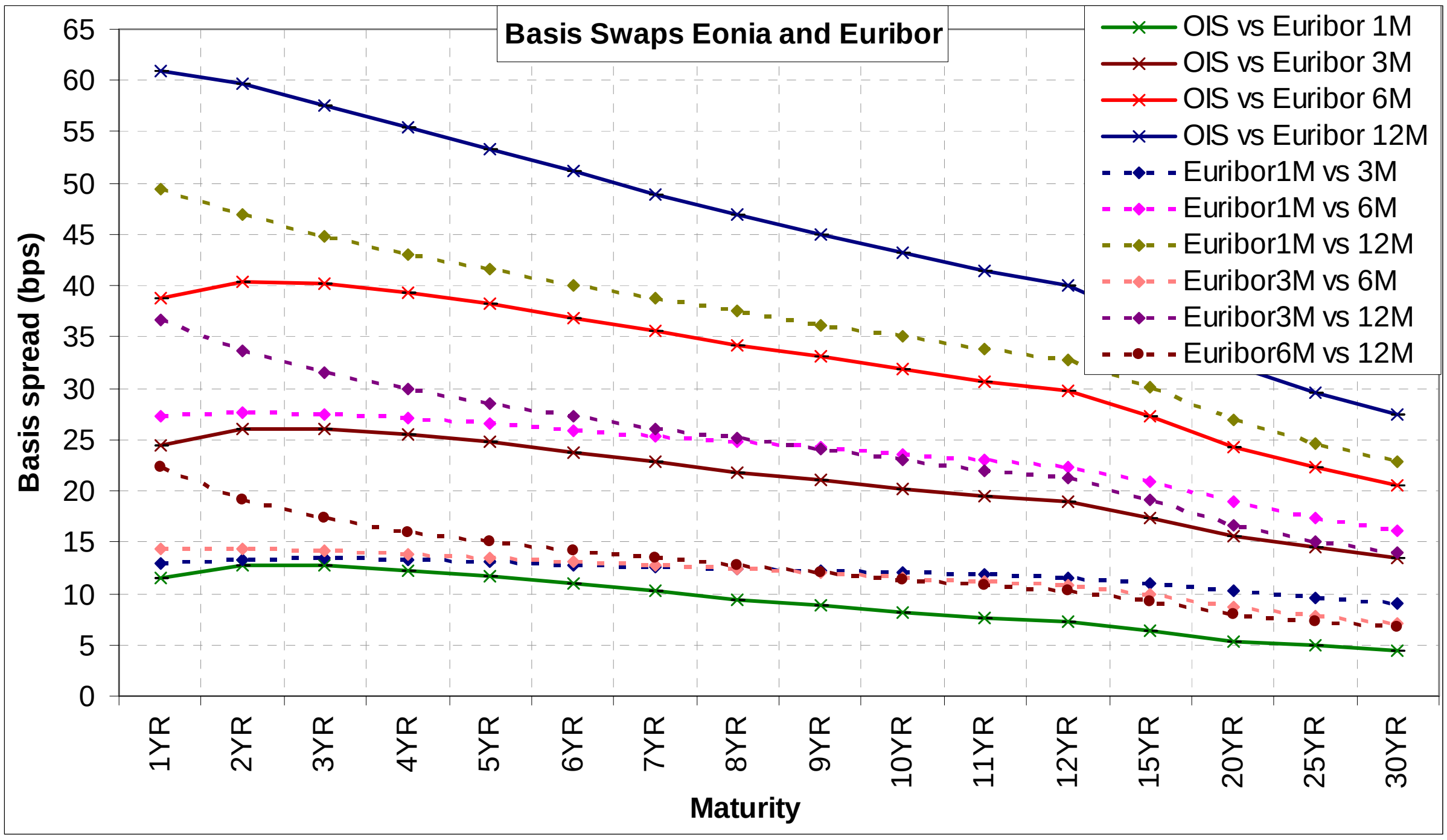

**Figure 2:** EUR Basis Swap term structures as of 30 June 2011.
Solid lines: Eonia-Euribor Basis Swaps spreads. Dashed lines: Euribor-Euribor Basis Swaps spread.

As a consequence, Libor/Euribor were seen as good market proxies for risk free rates, and used as the basic building block of no-arbitrage pricing theory, as described in pre-crisis textbooks on interest rate modelling (see e.g. [21], [22]). Following these assumptions, one easily obtained that all streams of floating Libor payments with equal maturity and different rate tenors, as those shown in Figure 3, had equivalent values and could be replicated one with each others.

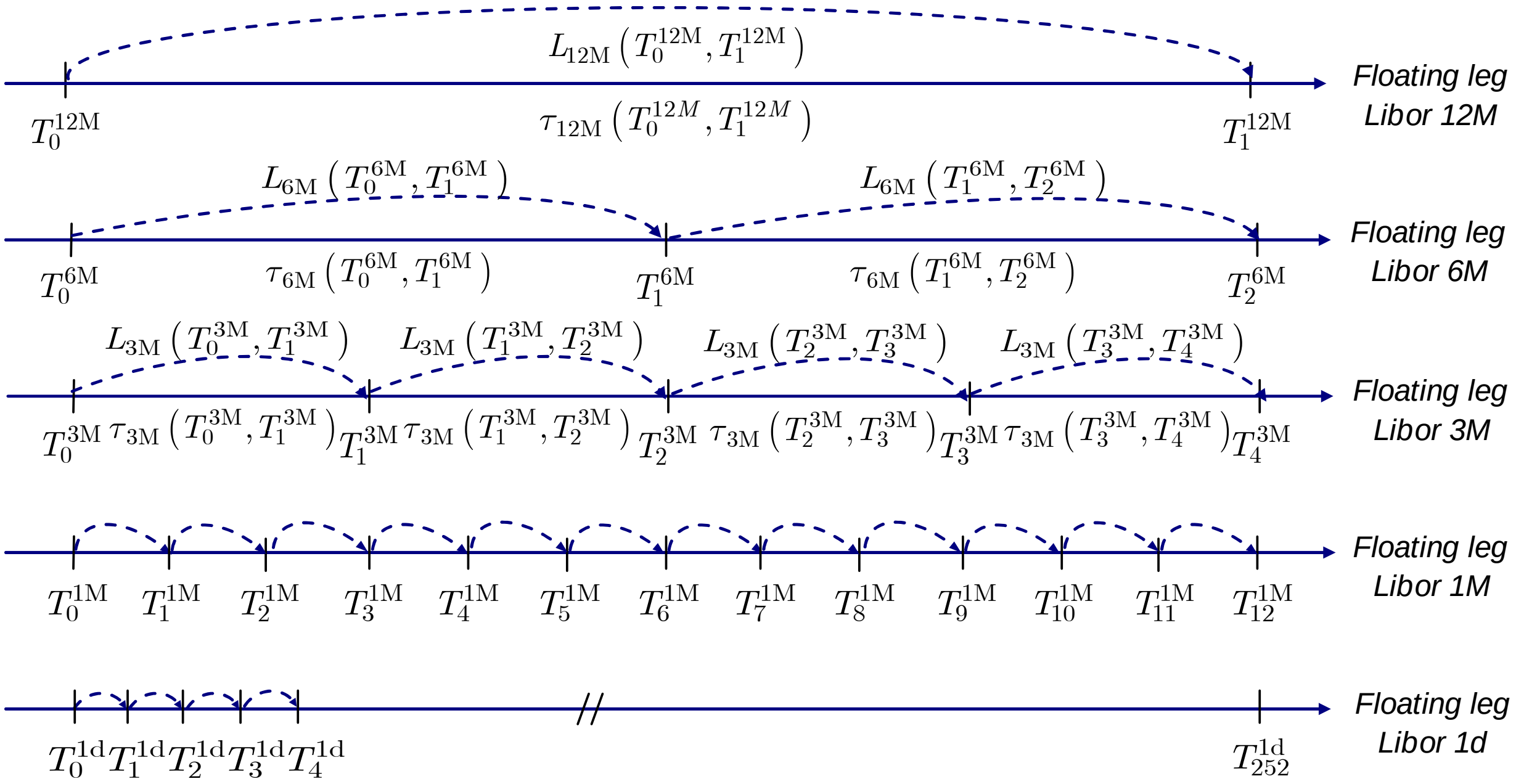

**Figure 3:** picture of floating Swap legs with equal maturities ($T_1^{12M} = T_2^{6M} = T_4^{3M} = T_{12}^{1M} = T_{252}^{1d} = 1Y$) and different Libor tenors (12M, 6M, 3M, 1M, 1d from top to bottom).

In mathematical terms, using notation and eq. (33) in Appendix B, one obtained the equalities given in the following eq. (1).

$$\begin{aligned}
\mathbf{Swap}(t;\mathbf{T}) &\rightarrow P(T_0;T_1^{12M})L(T_0^{12M},T_1^{12M})\tau(T_0^{12M},T_1^{12M}) \\
&\simeq \sum_{j=1}^{2} P(T_0;T_j^{6M})F(T_{j-1}^{6M},T_j^{6M})\tau(T_{j-1}^{6M},T_j^{6M}) \\
&\simeq \sum_{j=1}^{4} P(T_0;T_j^{3M})F(T_{j-1}^{3M},T_j^{3M})\tau(T_{j-1}^{3M},T_j^{3M}) \\
&\simeq \sum_{j=1}^{12} P(T_0;T_j^{1M})F(T_{j-1}^{1M},T_j^{1M})\tau(T_{j-1}^{1M},T_j^{1M}) \\
&\simeq \sum_{j=1}^{252} P(T_0;T_j^{1d})F(T_{j-1}^{1d},T_j^{1d})\tau(T_{j-1}^{1d},T_j^{1d}) \\
&\simeq 1 - P(T_0;T_1^{12M}), \\
F(T_{j-1}^{x},T_j^{x}) &= \frac{1}{\tau(T_{j-1}^{x},T_j^{x})}\left[\frac{P(T_0;T_{j-1}^{x})}{P(T_0;T_j^{x})} - 1\right],
\end{aligned} \tag{1}$$

As a consequence, the corresponding Basis Swaps, being constructed by two floating Libor legs with different tenors, displayed negligible basis spreads. Within this context, we call this effect *Libor tenor symmetry* or *replication invariance*.

The credit crunch crisis has begun precisely with the explosion of the credit and liquidity risk of primary banks, even those included in the Libor panels. The assumption of “too big to fail” has been definitely abandoned after the Lehman default. Thus, nowadays, after the crunch, banks are risky, interbank unsecured Deposits are risky as well, and the market value of this risk is embedded into Libor quotes. The longer the maturity of the Deposit contract (the Libor tenor), the higher the credit and liquidity risk embedded in the associated Libor rate. Hence floating Libor payments with equal maturity and different rate tenors as in Figure 3 have not longer equivalent values as in eq. (1), cannot be replicated with each other, and the corresponding Basis Swaps display huge basis spreads, as shown in Figure 1 and Figure 2 above. We call this effect *Libor tenor* or *replication symmetry breaking*.

Furthermore, also the market fixing mechanism makes an important difference. On one side, Libor is based, by definition, on banks’ perceptions of their own cost of funding in the OTC interbank market, not on actual transactions on an exchange market visible to other parties. Thus, in difficult times, when the money market may be very illiquid, Libor may be misrepresented, because just a few Libor contributors have actually traded in size at a given day, maturity and currency, and the others are forced to contribute a guess. Furthermore, Libor may be biased by those contributors not willing to show their funding troubles, as reported in [3] and argued e.g. in [4]. On the other hand, overnight rates, like Eonia (Euro OverNight Index Average) or US Federal fund effective rate, are based on volume-weighted averages of actual transactions during the day, and thus they reflect the true (over night) cost of funding realised on the market.

From the considerations above it is clear why overnight rates are nowadays the best available market proxy to risk-free rates. Thanks to their safe fixing mechanism and shortest possible tenor, they incorporate the smallest possible amount of counterparty default risk: precisely, the risk of default during the night (or the week end), given the market information available in the afternoon at close of business time. Hence eqs. (1) above must be modified, denoting with $P_d(t;T_j)$ the risk free Zero Coupon Bond for maturity $T_i$, and with $F(t;T_{j-1},T_j)$ the FRA rate fixed at $T_{j-1}$ and payed at $T_j$, thus obtaining that all the floating Libor payment streams differ from each other, as in the following eq. (2).

$$
\begin{aligned}
\mathbf{Swap}_{12\mathrm{M}}(t;\mathbf{T}) &= P_d(T_0;T_1)L_{12M}(T_0^{12M},T_1^{12M})\tau_{12M}(T_0^{12M},T_1^{12M}) \neq \\
\mathbf{Swap}_{6\mathrm{M}}(t;\mathbf{T}) &= \sum_{j=1}^{2} P_d(T_0;T_j)F_{6M}(T_{j-1}^{6M},T_j^{6M})\tau_{6M}(T_{j-1}^{6M},T_j^{6M}) \neq \\
\mathbf{Swap}_{3\mathrm{M}}(t;\mathbf{T}) &= \sum_{j=1}^{4} P_d(T_0;T_j)F_{3M}(T_{j-1}^{3M},T_j^{3M})\tau_{3M}(T_{j-1}^{3M},T_j^{3M}) \neq \\
\mathbf{Swap}_{1\mathrm{M}}(t;\mathbf{T}) &= \sum_{j=1}^{12} P_d(T_0;T_j)F_{1M}(T_{j-1}^{1M},T_j^{1M})\tau_{1M}(T_{j-1}^{1M},T_j^{1M}) \neq \\
\mathbf{Swap}_{1\mathrm{d}}(t;\mathbf{T}) &= \sum_{j=1}^{252} P_d(T_0;T_j)F_{1d}(T_{j-1}^{1d},T_j^{1d})\tau_{1d}(T_{j-1}^{1d},T_j^{1d}) \\
&\neq 1 - P_d(T_0;T_1^{12M}). \qquad (2)
\end{aligned}
$$

We refer to Appendix B for a brief summary of the quantities used above and to ref. [5] for a detailed explanation and a mathematical treatment of this problem.

We conclude our analogy by observing that the credit crunch has forced a *symmetry breaking* in the money market: from a “classical” situation characterized by negligible credit and liquidity risk in which “all the Libors were equal”, showing replication invariance under Libor tenor transformations and negligible basis spreads, towards a new “modern” market hierarchy (or segmentation) characterized by non-negligible credit and liquidity risk in which “some Libors are more equal than others”, showing broken Libor tenor symmetry, no replication invariance and large basis spreads. We stress that such Libor spectroscopy was already present and well known to market players before the credit crunch, as discussed e.g. in ref. [6], but not effective due to negligible basis spreads.

This is what we call the *Zeeman effect in finance*, as depicted in Figure 4 and summarized in Table 1, in which we show the one-to-one correspondences between physical and financial concepts.

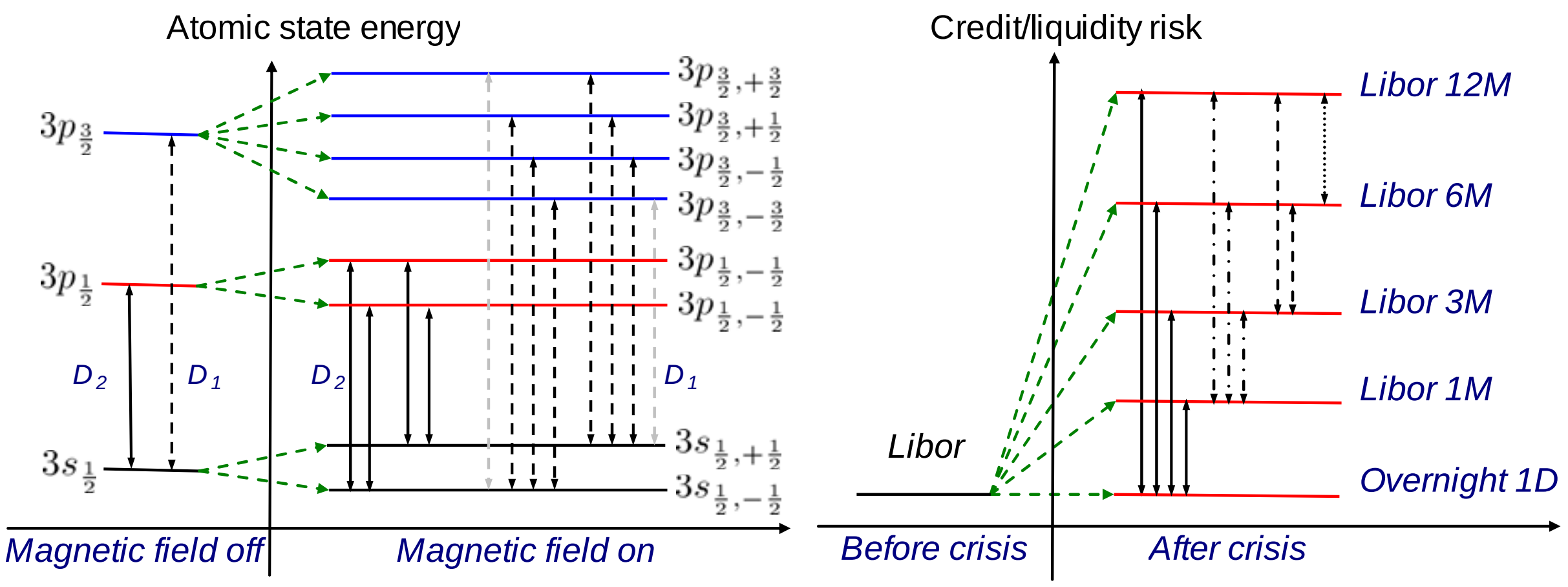

**Figure 4:** qualitative picture of the Zeeman effects in physics and finance. Left panel: under the magnetic field, three Sodium *sp* atomic states and two $D_1$ , $D_2$ spectral lines (left side) split into three multiplets of 2+2+4 states at different energies and two multiplets of 4+6 spectral lines (2 lines in gray in $D_1$ are not observable due to quantum suppression) at different wavelengths, respectively (right side). Right panel: under credit and liquidity risk, Libor (left side) splits into its components with different tenors (right side), and the corresponding Basis Swap spread diverge from zero. In particular, the Overnight 1D-Libor multiplet (4 full vertical lines) and the three Libor-Libor multiplets (3+2+1 dashed vertical lines) correspond exactly to the Eonia-Euribor and Euribor-Euribor Basis Swaps in Figure 2. More details in Appendices A and B.

| | **The Zeeman effect in Physics** | **The Zeeman effect in finance** |
|---|---|---|
| Discovery | Pieter Zeeman, Leyden, Aug. 1896. | Global markets, Aug. 2007. |
| Observation | Atomic spectral lines in a magnetic field split into groups, called Zeeman multiplets (Zeeman, 1896). | Basis Swaps with different tenors show, after the credit crunch, large gaps, called basis spreads. |
| Technology | Light spectroscopy, spectrometer. | Libor spectroscopy, global markets, market data providers. |
| Interpretation | In normal conditions (i.e. zero magnetic field):<br>o distinct atomic states displaying symmetry properties (e.g. rotational invariance) have the same energy;<br>o the corresponding spectral lines overlaps. | In pre-credit crunch conditions (i.e. negligible credit/liquidity risk),<br>o distinct Libor floating payments with equal maturity and different tenors have similar values (replication invariance);<br>o the corresponding Basis Swaps display negligible basis spread. |
| | The external magnetic field breaks the atomic rotational symmetry and splits the atomic states at different energies. | The credit and liquidity risk breaks the Libor tenor symmetry and splits equivalent Libor floating legs at different values. |
| | The corresponding spectral lines split into observable multiplets. | The corresponding Basis Swaps split into non-negligible basis spreads. |
| Differences | The magnetic field may be set by the experimenter at a fixed and constant magnitude. | The credit/liquidity risk is decided by the market and is intrinsically stochastic. |
| | The zero magnetic field limit with perfect rotational symmetry and no Zeeman splitting is possible and observable. | The zero credit/liquidity risk with perfect replication and zero basis spreads is an idealized case not observable in practice. |
| Consequences | Electron discovery, atomic structure, quantum mechanics. | Risky interest rates, multiple yield curves, CSA-discounting, multiple-curve pricing models. |

**Table 1:** similarities and differences between the Zeeman effects in physics and finance.

## 3. Consequences

The discovery of the Zeeman effect and its interpretation by Lorentz triggered deep and far consequences in physics [4], contributing to the development of quantum mechanics and, ultimately, to the modern science. Analogously, after the credit crunch, the market segmentation into risky Libors with huge Basis Swap spreads has induced deep and far consequences in finance, contributing to the development of a modern market in which credit and liquidity are recognized sources of risk and play a crucial role. In what follows we briefly introduce and discuss these news.

### 3.1. Market quotations

A first main consequence has been the large diffusion of Collateral Agreements (CSA, or “Credit Support Annex” in the ISDA standard documentation) as the main instrument to reduce the counterparty risk. According to the ISDA Margin Survey 2011 [7], most of the interbank counterparties and OTC derivatives’ transactions are (bilateral, cash) collateralised. This fact has the important consequence that one is allowed to look at derivatives’ prices quoted on the interbank market as referring to transactions under CSA. It is worth to stress at this point a frequent misunderstanding: thinking to interest rate derivatives (Swaps for instance), the CSA reduces the component of credit and liquidity risk associated to the specific counterparties involved in the financial contract, but it does not affect the component of credit and liquidity risk intrinsically carried by the underlying Libor rate, which is linked to the average credit and liquidity risk of the interbank money market, or, more precisely, of the Libor panel banks. Thus we arrive at the counter-intuitive, but correct, idea that Swaps under CSA are not counterparty risk free. The proof is essentially empiric: Basis Swap spreads quoted on the interbank market are both collateralised and far from zero at the same time. Not surprisingly, risks walk out of the door and return back from the window.

[4] By the way, Zeeman and Lorentz were also able to measure the ratio e/m between the electrical charge and the mass of the electron, one of the most important constants in physics, one year before the experimental discovery of the electron by J. J. Thomson in 1897.

### 3.2. Cost of funding and CSA-discounting

The CSA diffusion in the OTC market has led to a second very important consequence, called CSA-discounting. Prior to the crisis, there was a general consensus on using Libor to discount future cash flows generated by OTC derivatives, because Libor represented the average funding rate in the interbank market and was considered a good proxy to a risk free rate, as required by the standard no-arbitrage mathematical framework for pricing derivatives. Nowadays Libor is ruled out as discounting rate because it is risky, and even tenor dependent. Which Libor should be used ? Libor 6M ? Libor 3M ? Libor overnight ?

The answer is that, by no arbitrage, the discounting rate is the funding rate. Thus, in presence of CSA, the discounting and funding rate is precisely the collateral rate, which is normally overnight (e.g. Eonia, Sonia, Fed Funds, etc.), as implied by market OIS (Overnight Indexed Swaps). This is very welcome for no-arbitrage theory, because, as discussed above, overnight rates are presently the best proxy of risk free rates available on the market. Thus Libor has been abandoned as discounting rate in favour of the true funding rate set by the CSA.

In case of transactions without CSA, no-arbitrage still requires a discounting rate that reflects the bank's cost of funding on the market. The construction of such a *funding curve* is a difficult task because the bank has typically many sources of liquidity: the money and repo markets, the bond market, corporates' credit lines, retail bank accounts and mortgages' prepayment, etc. All these sources are characterized by their own interest rates, maturities and volumes that evolve over time in a non-deterministic (stochastic) way. Furthermore, the traditional money market funding may be just a minor fraction of the Bank's total funding capability, in particular for large international retail banks.

We stress that if each bank actualises future cash flows using its own market cost of funding, there is no more price agreement between counterparties. This is a question currently under theoretical investigation (see e.g. refs [8]-[12]). Further complications arise in case of non-standard CSA, e.g. one-way (asymmetrical) CSA, or in case the option to switch the collateral currency can be exercised ("CSA chaos", see e.g. ref. [13]).

### 3.3. Multiple curve world

A third consequence of the credit crunch is the multiple curve world. Prior to the crisis a single yield curve was sufficient for computing both discount and forward rates for market-coherent pricing of interest rate derivatives. Nowadays the same operation requires a multiplicity of yield curves. First, discount rates must be computed from (at least) two different curves: the OIS curve or the funding curve, depending weather the deal is under CSA or not. Then, FRA rates with different tenors must be computed from the corresponding tenor-dependent yield curves, constructed coherently from market quotes of instruments with the same tenor, and with the chosen discounting curve. Thus the classical exercise of yield curve construction (even discarded in some financial training) is nowadays at the heart of modern pricing theory and practice (see e.g. refs. [14]-[20]).

It is worth to stress that multiple curves construction implies the usage of multiple market data, often illiquid or even absent on the market. Furthermore, multiple curves implies also multiple delta sensitivities, or, in other words, multiple basis risk. Hence, hedging the risk of even plain vanilla interest rate derivatives (e.g. Swaps) become a very complex task in the post credit crunch world.

The transition to the multiple curve world has triggered a revival of theoretical research about multiple-curve pricing models for interest rate derivatives. Prior to the crisis a single fundamental stochastic variable, e.g. a short rate, and a single dynamical process (Hull-White, for instance) was sufficient to model the term structure of interest rates. Nowadays, the market segmentation implies that multiple correlated dynamical processes are required to model the multiple interest rate term structures observed on the market. Clearly this is not an easy task, and the research is at the very beginning (see for example ref. [20]). In particular, the calibration of such models to the market is very difficult, because part of the model parameters are related to non-quoted instruments (options on the basis, for instance). A technique frequently used by market practitioners is to freeze the dynamical evolution of the basis term structure at its present value (basis freezing). Clearly this is equivalent to discard the contribution of the basis volatility and correlations to the price of derivatives, an approximation that should be kept under control.

### 3.4. Switching to CSA-discounting in practice

A fourth, very important point is how to switch financial institutions to CSA-discounting in practice. This is not an easy task at all because of a variety of issues, discussed below.

#### *3.4.1. Market issues*

To date, some major clearing houses (e.g. LCH.Clearnet) have adopted the new methodology for interest rate Swaps. Some major broker (e.g. ICAP) has declared the adoption for interest rate derivatives (Swaps and options), and has changed the quotation pages consistently. The swaption market, in particular, has switched to forward premium quotation (as in the fx option market) precisely to reduce the impact of discounting. These changes can be occasionally checked on the market, in particular when in/out of the money interest rate derivatives are traded. Swaps, for instance, show a negligible sensitivity to the discount rate when their value is close to zero, as usual in standard trades at par, because the two legs have similar but opposite values and sensitivities. On the other side, the discount sensitivity may be important when the Swap is traded not at par (out of the money), e.g. in case of unwinding, because the netting between the two legs is only partial. Interest rate Cap/Floor/Swap options, instead, generally show smaller effects, because, once the market premium is fixed by supply and demand, the different discounting tends to be balanced by the different implicit volatility. The cancellation is perfect, by definition, for market options, and partial for off market options and legacy trades. Similar arguments apply to other market derivatives on different asset classes (inflation, fx, equity, credit), with the difference that the discounting effect may be further obscured by other sources of uncertainty, such as the inflation/fx curves/volatilities, the equity dividends and repo rates, the default probability curves, etc.

In conclusion, disentangling the effect of CSA-discounting into market or counterparty prices is not an easy task, especially for non-interest rate derivatives and for non-CSA counterparties.

#### *3.4.2. Collateral and liquidity issues*

Evidence of CSA-discounting from collateral management may be controversial. Besides the intrinsic difficulties cited above, collateral margination is usually managed by collateral desks at portfolio level for each counterparty under CSA, and not at trade level, thus hiding the discounting effects even more. On the other hand, in case of disputation pricing details are shared between the two counterparties in order to match the mark to market of the collateralised trades, thus allowing much more market intelligence than usual. Another bias may be introduced by opportunistic counterparties posting or asking collateral using the most convenient discounting methodology. Further complications arise because of the typical variety of clauses and details of collateral agreements, such as haircuts, margination frequency, rate spreads, currency, one-way margination, etc. that require, in principle, more sophisticated and CSA-dependent pricing methodologies. The new ISDA standardised CSA [18] should address and simplify these issues.

A very important challenge is the front-to-back integration of Banks' internal credit and funding management, from trading to treasury, collateral and back office, in order to benefit of centralised credit and liquidity charges at single trade level. Such a re-organisation of traditionally separated areas may result to be very difficult in large international banking groups characterised by an holding with multiple subsidiaries and locations. In particular, the yield curves used for pricing internal deals (trades between different legal entities inside the group) should reflect the true cost of internal funding within the group.

### 3.4.3. Accounting issues

International Accounting Standards (IAS) affirm that, "in determining the valuation of OTC derivative "a valuation technique (a) incorporates all factors that market participants would consider in setting a price and (b) is consistent with accepted economic methodologies for pricing financial instruments" (AG76). Thus there is a *judgemental area* where the estimation of fair value is based on market (multilateral) consensus. CSA-discounting is a typical case of evolution of the market consensus regarding the meaning of CSA: from a simple accessory guarantee to a determinant of the fair value.

Hedge accounting, in particular, is an accountancy practice allowed by IAS39 to mitigate the Profit & Loss volatility due to derivatives used for hedging. A typical situation arises when the interest rate risk of a liability (a bond issued by the bank for instance) is hedged using a Swap. Hedge accounting requires that the profit & loss of the package remains confined in the 80%-125% window with respect to the initial NPV. The pricing of the package is based on ad hoc methodologies (e.g. the liability cash flows are discounted using the floating rate of the Swap, for instance), that may partially account for the basis risk existing between the liability and the derivative. As a consequence the adoption of CSA-discounting may realize the basis risk, resulting in significant NPV jumps and even breaks of the hedge accounting 80-125 constrain. Hence, either the methodology must be revised to account for the basis risk, or hedges must be renegotiated.

### 3.4.4. IT issues

The adoption of CSA-discounting is a big issue from an IT point of view that requires huge resources to be properly addressed. Here are some critical points.

- Booking of trades in pricing systems must be reviewed such that the information regarding the collateral is recovered.
- Pricing systems configurations must be reviewed for CSA-compliance, allowing proper assignments to each trade of different yield curves depending on the CSA.
- Multiple yield curves and volatilities bootstrapping must be properly implemented and configured in all pricing systems.
- Price and risk computations must be reviewed in order to avoid hidden assumptions regarding discounting, e.g. automatic default yield curve usage without explicit assignment.
- Commercial systems require new releases able to manage CSA-discounting. Vendors must be typically fed with appropriate specs and the new releases carefully tested.
- Proprietary financial libraries must be reviewed and re-engineered to make them multiple-curve compliant. Previous poor library design is likely to require much more re-implementation effort.
- Systems integration and alignment must be carefully checked to avoid the classical "two systems two prices" problem.

In general, we can say that the switch to CSA-discounting is a kind of stress test for the IT architecture of a bank. The most confused IT situations typically imply much more effort to switch, and vice versa.

### *3.4.5. Risk Management issues*

The main risk management issues involved into CSA-discounting are below.

- **Market risk**: the most important source of market risk involved in CSA-discounting is the basis risk in the multiple-curve world, in which even plain vanilla interest rate derivatives (e.g. Swaps) display complex delta exposures distributed across multiple Libors with different tenors and OIS rates. This kind of risk may be not fully captured or represented in standard, old style pricing frameworks grounded on Libor discounting. Basis risk is also expensive to hedge, requiring market Swaps, OIS and Basis Swaps. In practice, it is often partially hedged in the classical fashion, using standard Libor Swaps, thus leaving an open exposure to the Libor-OIS basis. The latter may be huge (Fig. 4) and volatile (Fig.3). The corresponding (un)expected profit & loss is typically realized in case of unwindings or in case of adoption of CSA-discounting, for instance when trades are migrated to Central Counterparties.
- **Model risk**: this source of risk regards the usage of interest rate models (e.g. Hull-White, Libor Market Model, HJM, etc.) for pricing and hedging derivatives in a multiple-curve world. On the one hand, classical single-curve models imply strong approximations and do not capture the basis risk, that will be revealed once the models are updated in the form of an unexpected NPV jump. On the other hand, modern multiple-curve models may be able to give a better description of the basis risk but, to date, they are more complex, still under development, and there is no standard on the market.
- **Counterparty risk**: this source of risk is captured in CSA-discounting in the sense that, for trades under CSA, the collateral reduces the counterparty risk and the OIS-discounting ensures no-arbitrage between the collateral rate and the discounting rate. A residual source of counterparty risk is left behind by re-hypothecation issues and by the mechanics of margination [19]. In case of absence of CSA, Credit Value Adjustment (CVA) and Debt Value Adjustment (DVA) must be calculated. We stress that a consistent treatment of DVA and funding is an open topic still under investigation (see e.g. [9]-[11]).
- **Funding liquidity risk**: with funding liquidity risk we mean the risk of changing market funding rate. Funding liquidity risk management under CSA-discounting is complicated by the fact that derivatives have a funding impact that depends on the CSA. Hence a centralised liquidity management, integrating treasury, collateral management and sales/trading desks, would allow both a full view of all the expected cash flows generated by the bank's activity by derivatives in particular, and a correct pricing of funding costs at single trade level.
- **Operational risk**: the main source of operational risk (the risk of loss resulting from failed internal processes, people, systems, or external events) generated by CSA-discounting is related to the increasing complication of pricing systems and liquidity management discussed above. A typical example may be a wrong assignment between a deal or a group of deals and their CSA, resulting in a wrong pricing. An unexpected Profit & Loss is revealed when the mistake is fixed.

We conclude with the observation that the main driver of the switch to CSA-discounting is the evolution of pricing and risk methodologies, under the pressure of market evolution after the credit crunch. This a typical situation in which a solid Risk Management with strong quantitative resources may serve as the pivot of the innovation.

### *3.4.6. Management issues*

Management is called to lead the change, and the corresponding frictions, taking business opportunities and controlling risks and costs. The main management decisions required for switching to CSA-discounting, as discussed in the points above, regard:

- timing: when to switch
- how to switch: all together or piecewise, depending on currency, asset classes, desks, subsidiaries, time-zone, main trading markets, etc.

- a clear view about the multiple funding sources of the Bank (the funding curve) and re-organisation for centralised credit and liquidity management
- review and cleaning of collateral agreements with counterparties
- how to manage the basis risk and the Profit & Loss generated by the switch
- how to manage the hedge accounting
- IT upgrade: booking, pricing, reporting, etc.
- communication and explanation of the switch to markets, customers, auditors and regulators.

#### *3.4.7. The role of Quants*

It is clear from the discussion above that CSA-discounting is a typical complex problem in which a simple no-arbitrage pricing issue (choosing the correct discounting curve) generates many consequences that propagate all around in the market and inside the banks. In such a situation quant people have the responsibility of extending the modern no-arbitrage pricing framework into other areas of the bank, traditionally not familiar with pricing issues, in order to reach a better fair value and risk management at Bank's level.

## 4. Conclusions

Once upon a time...
there was a classical financial world in which all the Libors were equal. Nowadays credit, liquidity and basis risk plays a crucial role. The Interbank market has developed an high degree of collateralisation, such that we may look at derivatives' quotations as referring to transactions under CSA. The cost of funding OTC derivatives has become a central topic in derivatives pricing. Multiple yield curves must be bootstrapped from multiple market data, and even plain vanilla interest rate Swaps display complex risk exposures distributed across multiple Libors and funding rates with different tenors. Multiple-curve interest rate models must be developed for pricing both plain vanilla and exotic derivatives. Switching to CSA discounting is not only a pricing issue of changing discount factors, but implies multiple issues investing all areas of the Bank, and a change of paradigm towards a more integrated management of Bank's funding across trading, treasury and collateral.
... and they all lived happily ever after ...?

## Appendix A: Quantum mechanics and Zeeman effect in a nutshell

In this Appendix we report a short description of the concepts and quantities cited in section 1 above. For more details see e.g. ref. [2].

The physics of atoms and molecules is described by one of the best scientific theories ever conceived by mankind: *quantum mechanics*. It is based on the experimentally observed dual particle-wave nature of matter and energy, and it is formalised into a set of postulates, known as the *Copenhagen interpretation*, prescribing the mathematical rules to describe a quantum system and to compute its physical properties. In a nutshell, any physical system $\mathcal{C}$ is associated with an appropriate *Hilbert space* $\mathcal{H}_{\mathcal{C}}$. Any physical state of $\mathcal{C}$ is described by an appropriate element $\Psi(\mathbf{x},t) \in \mathcal{H}_{\mathcal{C}}$, where ***x*** is a generic vector containing all the fundamental variables (e.g. positions of the system's particles in the 3D space), called *wavefunction* or *state vector*, that contains all the possible informations on $\mathcal{C}$. The dynamics of the state is governed by the *Schrödinger equation*

$$\hat{H}\Psi(\mathbf{x},t) = i\hbar\frac{\partial}{\partial t}\Psi(\mathbf{x},t),$$
$$\Psi(\mathbf{x},0) = \Psi_0(\mathbf{x}),$$
$$\langle\Psi(t)|\Psi(t)\rangle := \int |\Psi(\mathbf{x},t)|^2\, d\mathbf{x} = 1, \tag{3}$$

where $\hat{H}$ is the Hamiltonian operator representing the total (kinetic + potential) energy of the system, $\hbar = 1.054571726(47)\times 10^{-34}$ Joule x second is the reduced Planck's constant, and *i* is the imaginary unit. The time-independent version of the Schrödinger equation, suitable for stationary systems (like atoms), is

$$\hat{H}\Psi(\mathbf{x}) = E\Psi(\mathbf{x}), \tag{4}$$

where *E* is the energy (eigenvalue of $\hat{H}$) of the state $\Psi(\mathbf{x})$ (eigenvector of $\hat{H}$).

Any physically observable quantity *Q* (e.g. energy, position, momentum, magnetic field, etc.) is considered as a stochastic variable associated with a probability measure and with an operator $\hat{Q}$ in $\mathcal{H}_C$ , whose eigenvalues and eigenvectors

$$\hat{Q}\varphi_n = \alpha_n\varphi_n, \quad n = 1, ..., \tag{5}$$

constitute, under appropriate mathematical conditions, the spectrum of any possible value for *Q* and an orthonormal basis in $\mathcal{H}_C$, respectively. The probability that an observation of *Q* at time *t* provides the value $\alpha_n$ is given by the *Born rule*

$$P\left(Q = \alpha_n, t\right) = \left|\langle\varphi_n|\Psi(t)\rangle\right|^2 = \left|\int \varphi_n^*(\mathbf{x})\Psi(\mathbf{x},t)\, d\mathbf{x}\right|^2, \tag{6}$$

while the expected value and the variance of *Q* at time *t* are given by

$$\left\langle\hat{Q}\right\rangle_t := \left\langle\Psi(t)\left|\hat{Q}\right|\Psi(t)\right\rangle = \int_{\mathcal{H}}\Psi^*(\mathbf{x},t)\hat{Q}\Psi(\mathbf{x},t)\, d\mathbf{x} = \sum_n \alpha_n P(Q = \alpha_n, t),$$
$$(Q)_t := \left\langle\left[\hat{Q} - \left\langle\hat{Q}\right\rangle_t\right]^2\right\rangle_t = \sum_n \left[\alpha_n - \left\langle\hat{Q}\right\rangle_t\right]^2 P(Q = \alpha_n, t). \tag{7}$$

Given any two physically observables quantities $Q_1,\, Q_2$, the *Jordan's theorem*

$$\mathrm{Var}_t(Q_1)\mathrm{Var}_t(Q_2) \geq \frac{1}{2}\left|\left\langle\left[\hat{Q}_1, \hat{Q}_2\right]\right\rangle_t\right|,$$
$$\left[\hat{Q}_1, \hat{Q}_2\right] := \hat{Q}_1\hat{Q}_2 - \hat{Q}_2\hat{Q}_1, \tag{8}$$

holds, where $\left[\hat{Q}_1, \hat{Q}_2\right]$ is called *commutator*. If $\left[\hat{Q}_1, \hat{Q}_2\right] \neq 0$ then $Q_1,\, Q_2$ cannot be measured together with any precision and are called *incompatible observables*. For example, position *x* and momentum *p* of a one dimensional single free particle are incompatible because

$$\mathrm{Var}(x)\mathrm{Var}(p) \geq \frac{\hbar}{2}. \tag{9}$$

This is the famous *Heisenberg's uncertainty principle*. On the other side if $Q_1,\, Q_2$ commute, $\left[\hat{Q}_1; \hat{Q}_2\right] = 0$, they are said to be *compatible observables*, and can be physically measured together with any precision. In particular, any observable $\hat{Q}$ such that $\left[\hat{Q}, \hat{H}\right] = 0$ is said a *constant of motion* because

$$\frac{d}{dt}\left\langle\hat{Q}\right\rangle_t = \frac{1}{i\hbar}\left\langle\left[\hat{Q}, \hat{H}\right]\right\rangle_t = 0. \tag{10}$$

Moreover, $\hat{Q}_1, \hat{Q}_2$ commute if and only if they have a common orthonormal basis of eigenvectors

$$\hat{Q}_1\varphi_{n,m} = \alpha_n\varphi_{n,m},$$
$$\hat{Q}_2\varphi_{n,m} = \beta_m\varphi_{n,m}. \tag{11}$$

There are further postulates that incorporate into the theory all the experimental observations, e.g. the *Pauli exclusion principle*, the *correspondence principle* with classical mechanics, etc. Thus all that one must do to compute the physical properties of quantum system is to:

- o characterise the Hilbert space $\mathcal{H}_\mathcal{C}$ and the physical operators $\hat{Q}$ associated with the system $\mathcal{C}$;
- o model the Hamiltonian $\hat{H}$, describing the energy of the system $\mathcal{C}$;
- o choose the appropriate constants of motion and find the complete orthonormal basis in common with $\hat{H}$;
- o find the solution $\Psi(\mathbf{x}, t)$ of the time-dependent Schrödinger equation (3) as linear combination of the eigenvectors of $\hat{H}$;
- o compute the physically observable properties *Q* of $\mathcal{C}$ using eq. (6).

We may now apply the formalism above to the case of the Zeeman effect. The Hamiltonian for an atom in a magnetic field is given by

$$\hat{H} = \hat{H}_0 + \hat{V},$$
$$\hat{V} = -\hat{\boldsymbol{\mu}} \cdot \hat{\boldsymbol{B}},$$
$$\hat{\boldsymbol{\mu}} = -\frac{\mu_B}{\hbar}\left(g_L\hat{\boldsymbol{L}} + g_S\hat{\boldsymbol{S}}\right) = -\frac{\mu_B}{\hbar}\left[\hat{\boldsymbol{J}} + (g_S - 1)\hat{\boldsymbol{S}}\right],$$
$$\hat{\boldsymbol{J}} = \hat{\boldsymbol{L}} + \hat{\boldsymbol{S}},$$
$$\mu_B = \frac{q_e\hbar}{2m_e} = 9.27400915 \times 10^{-24} \text{ Joule x Tesla}^{-1},$$
$$g_L = 1,\ g_S = 2.0023192, \tag{12}$$

where $\hat{H}_0$ is the unperturbed Hamiltonian of the atom without magnetic field, $\hat{V}$ is the perturbation of the magnetic field $\hat{\boldsymbol{B}}$, $\hat{\boldsymbol{\mu}}$ is the magnetic moment operator of the atom, $\hat{\boldsymbol{L}}, \hat{\boldsymbol{S}}, \hat{\boldsymbol{J}}$ are the orbital, spin and total electronic angular momentum operators, respectively, $\mu_B$ is the *Bohr's magneton* (representing the minimum dipole magnetic moment of an atomic electron), and $g_l, g_s$ are the *L, S gyromagnetic ratios*, respectively. Taking a magnetic field constant over the atomic dimension and choosing our z axis parallel to $\hat{\boldsymbol{B}}$ we have

$$\hat{\boldsymbol{B}} = [0, 0, B_0],$$
$$\hat{V} = \frac{\mu_B B_0}{\hbar}\left[\hat{J}_z + (g_S - 1)\hat{S}_z\right],$$
$$\left[\hat{H}, \hat{\boldsymbol{L}}^2\right] = \left[\hat{H}, \hat{\boldsymbol{J}}^2\right] = 0, \left[\hat{H}, \hat{J}_z\right] = 0. \tag{13}$$

Denoting the unperturbed spectrum common to $\hat{H}_0, \hat{\boldsymbol{L}}^2, \hat{\boldsymbol{J}}^2, \hat{J}_z$ with

$$\hat{H}_0\varphi^0_{nljm_j} = E^0_{nljm_j}\varphi^0_{nljm_j},$$
$$\hat{\boldsymbol{L}}^2\varphi^0_{nljm_j} = l(l+1)\hbar^2\varphi^0_{nljm_j},$$
$$\hat{\boldsymbol{J}}^2\varphi^0_{nljm_j} = j(j+1)\hbar^2\varphi^0_{nljm_j},$$
$$\hat{J}_z\varphi^0_{nljm_j} = m_j\hbar\varphi^0_{nljm_j},\ m_j = -j, ..., j, \tag{14}$$

where $n$, $l$, $j$, $m_j$ are the quantum numbers associated with the four physical operators in eq. (14), one obtains, using perturbation theory, the total spectrum common to $\hat{H}$, $\hat{\boldsymbol{L}}^2$, $\hat{\boldsymbol{J}}^2$, $\hat{J}_z$ as

$$\begin{aligned}
\hat{H}\varphi_{nljm_j} &= E_{nljm_j}\varphi_{nljm_j}, \\
\varphi_{nljm_j} &= \varphi^0_{nljm_j}, \\
E_{nljm_j} &= E^0_{nljm_j} + E^1_{nljm_j}, \\
E^1_{nljm_j} &= \left\langle \varphi^0_{nljm_j} \left| \hat{V} \right| \varphi^0_{nljm_j} \right\rangle \\
&= \frac{\mu_B B_0}{\hbar} \left\langle \varphi^0_{nljm_j} \left| \hat{J}_z + (g_S - 1)\hat{S}_z \right| \varphi^0_{nljm_j} \right\rangle \\
&= \mu_B B_0 m_j g_J, \\
g_J &:= 1 + (g_S - 1)\frac{j(j+1) - l(l+1) + s(s+1)}{2j(j+1)},
\end{aligned} \tag{15}$$

where $g_J$ is called *Landé g-factor*. This is the final theoretical formula for the energy of electronic states in a magnetic field. We see that for non-null magnetic field $B_0$ the $m_j$ degeneracy – due to the conservation of angular momentum $J$ and $J_z$ by eq. (13) - is broken, and there is an energy splitting, equal to the term $E^1_{nljm_j}$, of the original atomic states with energy $E^0_{nljm_j}$ depending linearly on $B_0$ and on the quantum numbers associated with the states. In particular, any electronic state with given $n$, $l$, $j$ numbers is split into $2j+1$ states corresponding to $m_j$ values.

The result above apply to any atomic system. In particular, we may explain the Zeeman effect in Sodium (Figure ). The atomic structure of Sodium, similarly to the other alkaline metals, is characterised by a core of tightly bounded atomic electrons plus one single electron occupying the outermost, less bounded, electronic state. The two $D_1$ and $D_2$ spectral lines of Sodium correspond to transitions of this single electron between the ground state and the first two excited states,

$$\begin{aligned}
D_1: &\quad 3s_{\frac{1}{2}} \longleftrightarrow 3p_{\frac{1}{2}}, \quad \lambda_{D_1} = 589.59\, nm, \\
D_2: &\quad 3s_{\frac{1}{2}} \longleftrightarrow 3p_{\frac{3}{2}}, \quad \lambda_{D_2} = 589.00\, nm,
\end{aligned} \tag{16}$$

where we have used the standard spectroscopic notation for atomic states

$$\begin{aligned}
3s_{\frac{1}{2}} &:= \varphi^0_{3,0,\frac{1}{2},\pm\frac{1}{2}}, \\
3p_{\frac{1}{2}} &:= \varphi^0_{3,1,\frac{1}{2},\pm\frac{1}{2}}, \\
3p_{\frac{3}{2}} &:= \varphi^0_{3,1,\frac{3}{2},(\pm\frac{1}{2},\pm\frac{3}{2})}.
\end{aligned} \tag{17}$$

The magnetic field breaks the $3s_{\frac{1}{2}}$, $3p_{\frac{1}{2}}$ states into two states each, corresponding to $m_j = \pm\frac{1}{2}$, and the $3p_{\frac{3}{2}}$ state into four states, corresponding to $m_j = \pm\frac{1}{2}, \pm\frac{3}{2}$. Hence the two $D_1$ , $D_2$ spectral lines split into two multiplets of lines, differing from the mother lines of wavelengths given by

$$\begin{aligned}
\frac{1}{\lambda_{nljm_j,n'l'j'm'_j}} &= \frac{1}{hc}\left| E_{n'l'j'm'_j} - E_{nljm_j} \right| \\
&= \frac{\mu_B}{hc} B_0 \left| m'_j g'_J - m_j g_J \right|.
\end{aligned} \tag{18}$$

There are in principle 4 + 8 spectral lines, but only 4 + 6 lines are observable, as shown in Figure 4, according with the electromagnetic selection rules

$$\begin{aligned}
\Delta l &= \pm 1, \\
\Delta j &= 0, \pm 1, \\
\Delta m_j &= 0, \pm 1,
\end{aligned} \tag{19}$$

that suppress the two lines $3s_{\frac{1}{2},-\frac{1}{2}} \longleftrightarrow 3p_{\frac{1}{2},\frac{3}{2}}$, $3s_{\frac{1}{2},\frac{1}{2}} \longleftrightarrow 3p_{\frac{1}{2},-\frac{3}{2}}$ with $\Delta m_j = |m_j - m_j'| = 2$, in perfect agreement with the experimental evidence in Figure .

## Appendix B: No-arbitrage pricing and interest rate derivatives in a nutshell

In this Appendix we report a short description of the concepts and quantities cited in sections 2-3 above. For more details see e.g. refs. [5], [16]-[22].

Nowadays the most important fundamental variables for interest rate derivatives are the basic interest rates underlying these contracts, Libor, Euribor and Eonia in particular. **Libor** (London Interbank Offered Rate), sponsored by the British Bankers Association (BBA) and first published in 1986, is the reference rate for money market and over-the-counter (OTC) swap transactions. Libor is exchanged on the London money market through unsecured loans called Certificates of Deposit (see below) for 15 different maturities (or Libor tenors, from 1 day to 12 months), and for all the major world's currencies. Every business day it is fixed as the trimmed average (excluding the highest and the lowest 25%) of the contributions of a panel of primary dealers, answering to the question "At what rate could you borrow funds, were you to do so by asking for and then accepting inter-bank offers in a reasonable market size just prior to 11 a.m. London time?". The Contribution Panels are composed by 8-12-16 banks per currency, selected according to scale of market activity and reputation, reviewed twice a year. Thus Libor reflects the average perceived cost of funding of banks in the London interbank money market for each given tenor and currency.

**Euribor** (Euro Interbank Offered Rate), sponsored by the European Banking Federation (EBF) and first published on 30 December 1998, is the reference rate for Euro money market and OTC swap transactions. It is defined as "the rate at which Euro interbank Deposits are being offered within the EMU zone by one prime bank to another at 11:00 a.m. Brussels time". The rate fixing mechanics is similar to Libor, except for the averaging (the highest and lowest 15% are excluded) and for the panel, that gathers more than 40 banks selected according to their business volume and reputation in the Euro zone money markets, plus some large international bank from non-EU countries with important euro zone operations. Thus Euribor reflects the average perceived cost of funding of banks in the Euro interbank money market for each given tenor.

**Eonia** (Euro OverNight Index Average rate), sponsored by the EBF and first published on 4 January 1999, is the reference rate for Euro OTC overnight Deposits. Eonia is fixed every business day as the average rate (without exclusions) of the overnight transactions (one day tenor) executed during that day by the banks in the Euribor panel. Eonia is used by the European Central Bank (ECB) as a tool of effecting and observing the transmission of its monetary policy actions. Thus Eonia rate reflects the actual overnight cost of liquidity expectations of banks and the monetary policy effects in the Euro money market. Similar discussions apply to overnight rates in other currencies, such as Fed Fund rate (USD), Sonia (Sterling OverNight Index Average), etc.

The **no-arbitrage pricing framework** of financial derivatives is based on the fundamental pricing theorem: given a generic financial instrument $\Pi$ and a reference asset $N$, called numeraire, with payoffs $\Pi(T), N(T)$ at time $T$, respectively, under appropriate conditions there exist a unique martingale probability measure $Q_N$ associated to $N$, such that

$$\frac{\Pi(t)}{N(t)} = \mathbb{E}_t^{Q_N}\left[\frac{\Pi(T)}{N(T)}\right], \quad \forall t \leq T, \tag{20}$$

where $\Pi(t), N(t)$ are the prices of $\Pi$ and $N$ at time $t<T$ and $\mathbb{E}_t^{Q_N}[X]$ denotes the expectation of a stochastic variable X at time $T$ conditioned to the information set available at time $t$ under the probability measure $Q_N$. Two measures are particularly useful for pricing derivatives, because they are associated to the two simplest numeraires: the risk free Bank Account and the Zero Coupon Bond. The **Bank Account** is an ideal financial instrument representing an abstract loan that rewards its holder with the risk free rate, such that

$$
\begin{aligned}
B_d(0) &= 1, \\
dB_d(t) &= r_d(t)B_d(t)dt, \\
B_d(T) &= \exp\int_0^T r_d(t)\,dt, \\
D_d(t,T) := \frac{B_d(t)}{B_d(T)} &= \exp\left[-\int_t^T r_d(u)\,du\right], \\
\pi(t) &= \mathbb{E}_t^{Q_{B_d}}\left[D_d(t,T)\pi(T)\right],
\end{aligned}
\tag{21}
$$

where $r_d(t)$ is a risk free spot instantaneous *short rate* over the time interval $[t;t+dt]$, $D_d(t,T)$ is the risk free *stochastic discount* factor and $Q_{B_d}$ is the risk neutral measure associated to $B_d$. Thus the price $\Pi(t)$ at time $t$ of any derivative is the expectation of the discounted payoff $\Pi(T)$ at time $T>t$ under the risk neutral measure (last eq. above). The risk free Zero Coupon Bond is a contract in which one party guarantees to the other party the payment of one unit of currency at maturity date $T$, with no other payments, such that

$$
\begin{aligned}
P_d(T;T) &= 1, \\
P_d(t;T) = \mathbb{E}_t^{Q_{B_d}}\left[D_d(t,T)P_d(T;T)\right] &= \mathbb{E}_t^{Q_{B_d}}\left[D_d(t,T)\right], \\
\Pi(t) = P_d(t;T)\mathbb{E}_t^{Q_d^T}\left[\Pi(T)\right], &\quad \forall\, t \le T,
\end{aligned}
\tag{22}
$$

where $Q_d^T$ is the T-forward measure associated to $P_d(t;T)$. Thus the price at time $t$ of the risk free zero coupon bond is the expectation of the stochastic discount factor at time $T>t$ under the risk neutral measure (second line above), and the derivative's price $\Pi(t)$ can be written as the expectation of the payoff discounted with $P_d(t;T)$.

For interest rates, in particular, the **FRA rate** $F_{x,j}(t)$ associated to the Libor with tenor $x$ plays a central role. It is defined as the expectation at time $t$ of the Libor fixing with tenor $x$ at time $T_{i-1} > t, \quad i = 1, ..., N,$

$$
F_{x,i}(t) := \mathbb{E}_t^{Q_d^{T_i}}\left[L_x(T_{i-1},T_i)\right],
\tag{23}
$$

under the $T_i$-forward probability measure $Q_d^{T_i}$ associated to the Zero Coupon Bond $P_d(t;T_i)$, has the following interesting properties:

1. At fixing date $T_{i-1}$ it coincides with the Libor rate

$$
F_{x,i}(T_{i-1}) = L_x(T_{i-1},T_i).
\tag{24}
$$

2. At any time $t < T_{i-1}$ it is a martingale under the probability measure $Q_d^{T_i}$ such that

$$
F_{x,i}(t) := \mathbb{E}_t^{Q_d^{T_i}}\left[L_x(T_{i-1},T_i)\right] = \mathbb{E}_t^{Q_d^{T_i}}\left[F_{x,i}(T_{i-1})\right],
\tag{25}
$$

following the stochastic differential equation

$$
dF_{x,i}(t) = F_{x,i}(t)\sigma_{x,i}(t)dW_{x,i}^{Q_d^{T_i}}(t),
\tag{26}
$$

where $W(t)$ is a standard Brownian motion with zero mean and unit variance and $\sigma_{x,i}(t)$ is the FRA rate (lognormal) instantaneous volatility, such that

$$
\begin{aligned}
\mathbb{E}_t^{Q_d^{T_i}}\left[dF_{x,i}(T_{i-1})\right] &= 0, \\
\mathrm{Var}_t^{Q_d^{T_i}}\left[dF_{x,i}(T_{i-1})\right] &= \sigma_{x,i}^2(t)dt.
\end{aligned} \tag{27}
$$

3. In the limit of negligible counterparty/liquidity risk (no Libor segmentation) it recovers the classical single-curve value

$$
\begin{aligned}
F_{x,i}(t) &:= \mathbb{E}_t^{Q_d^{T_i}}\left[L_x(T_{i-1},T_i)\right] \to \mathbb{E}_t^{Q^{T_i}}\left[L(T_{i-1},T_i)\right] \\
&= \mathbb{E}_t^{Q^{T_i}}\left[F(T_{i-1};T_{i-1},T_i)\right] := F_i(t) \\
&= \frac{1}{\tau(T_{i-1},T_i)}\left[\frac{P(t;T_{i-1})}{P(t;T_i)} - 1\right].
\end{aligned} \tag{28}
$$

We will use the results above to derive the pricing formulas for the simplest interest rate derivatives.

Interest rate **Certificates of Deposit**, or simply Deposits (Figure 5), are standard zero coupon unsecured loans, exchanged on the money market between banks and major financial institutions, such that, at start date $T_0$, counterparty A (the Lender) pays a nominal amount $N$ to counterparty B (the Borrower) and at maturity date $T_i$ ($i=1,\dots,N$) the Borrower, if not defaulted, pays back to the Lender the nominal amount $N$ plus the interest accrued over the time interval $[T_0;T_i]$ (called Deposit or rate *tenor* $x$) at the annual simply compounded Libor rate $L_x(T_0,T_i)$ (or Euribor, Eonia, or any other rate), fixed just prior $T_0$ (two working days in the EUR market).

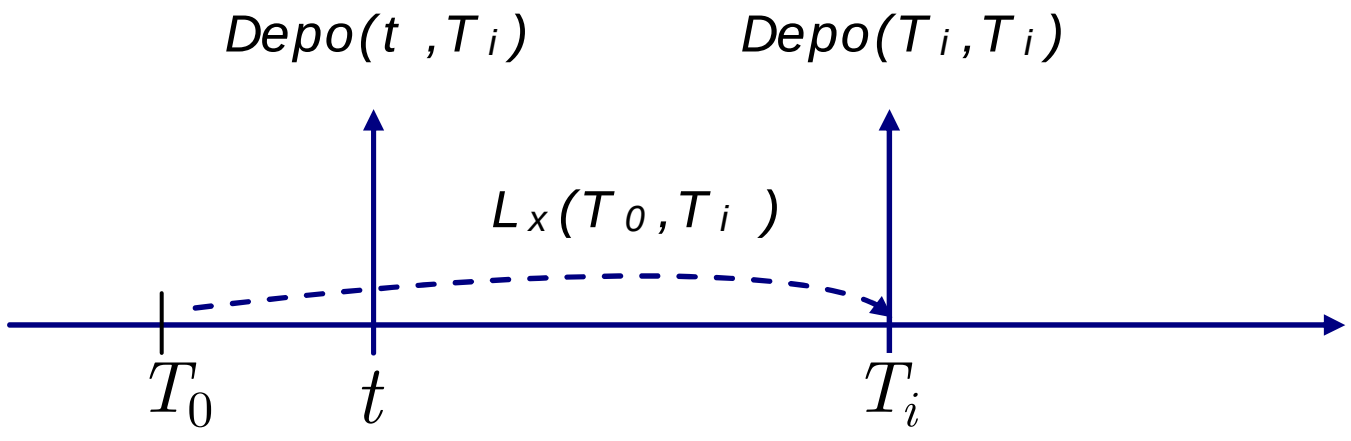

**Figure 5:** schematic picture of a Deposit contract.

Thus the payoff and the price of the Deposit at times $T_i$ and $t$ such that $T_0 < t < T_i$ (when the Libor is already known), respectively, are given by

$$
\begin{aligned}
\mathbf{Depo}(T_i;\mathbf{T}) &= N\left[1 + L_x(T_0,T_i)\tau_x(T_0,T_i)\right], \\
\mathbf{Depo}(t;\mathbf{T}) &= NP_d(t;T_i)\left[1 + L_x(T_0,T_i)\tau_x(T_0,T_i)\right],
\end{aligned} \tag{29}
$$

where $\tau_x(T_0,T_i)$ is the year fraction over $[T_0;T_i]$, with the appropriate day count convention, and we have assumed the existence of a risk free Zero Coupon Bond with payoff $P_d(T_i;T_i)=1$ at maturity $T_i$ and price $P_d(t;T_i)$ at time $t < T_i$ acting as discount factor.

Interest rate **Swaps** (Figure 6) are OTC contracts in which two counterparties agree to exchange two streams of cash flows, typically tied to a fixed rate K against floating rate $L_x$ with tenor $x$. These payment streams are called fixed and floating leg of the Swap, respectively, and they are characterized by two schedules S, T and coupon payoffs

$$
\begin{aligned}
&\mathbf{S} = \{S_0,\dots,S_n\}, \text{ fixed leg schedule}, \\
&\mathbf{T} = \{T_0,\dots,T_m\}, \text{ floating leg schedule}, \\
&S_0 = T_0, S_n = T_m, \\
&\mathbf{Swaplet}_{\text{fix}}(S_i;S_{i-1},S_i,K) = NK\tau_K(S_{i-1},S_i), \\
&\mathbf{Swaplet}_{\text{float}}(T_j;T_{j-1},T_j) = NL_x(T_{j-1},T_j)\tau_x(T_{j-1},T_j),
\end{aligned} \tag{30}
$$

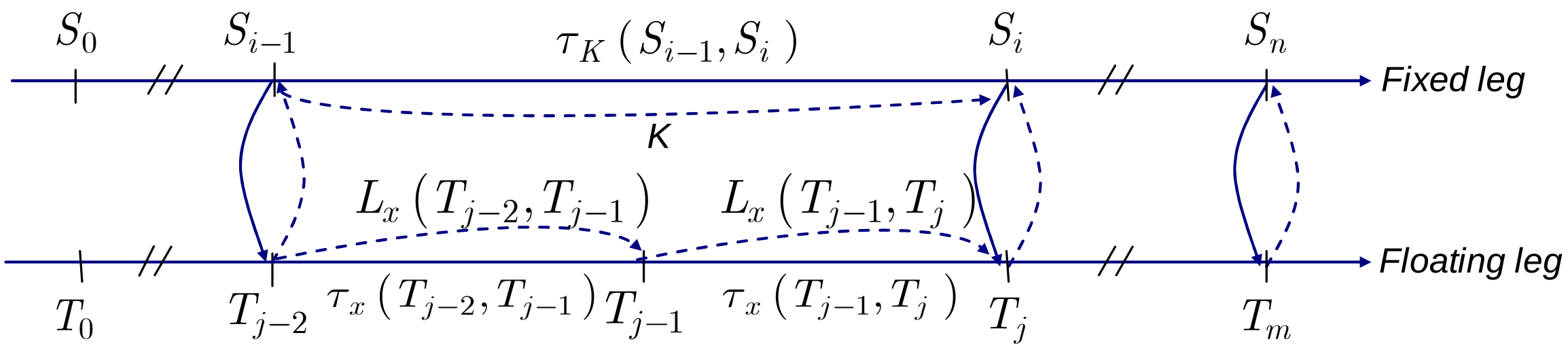

**Figure 6:** schematic picture of a fixed vs floating Swap.
Vertical full/dashed lines denote fixed/floating cash flows.

where $\tau_K$ and $\tau_x$ are the year fractions with the fixed and floating rate conventions, respectively. The price of the Swap, using eq. (23), is given by

$$\begin{aligned}
\mathbf{Swap}(t;\mathbf{T},\mathbf{S},K,\omega) &= \omega\left[\mathbf{Swap}_{\text{float}}(t;\mathbf{T}) - \mathbf{Swap}_{\text{fix}}(t;\mathbf{S},K)\right] \\
&= N\omega\left[\sum_{j=1}^{m} P_d(t,T_j)F_{x,j}(t)\tau_x(T_{j-1},T_j) - KA_d(t,\mathbf{S})\right] \\
&= N\omega\left[R_x^{\text{Swap}}(t;\mathbf{T},\mathbf{S}) - K\right]A_d(t,\mathbf{S}), \\
R_x^{\text{Swap}}(t;\mathbf{T},\mathbf{S}) &:= \frac{\sum_{j=1}^{m} P_d(t,T_j)F_{x,j}(t)\tau_x(T_{j-1},T_j)}{A_d(t,\mathbf{S})}, \\
A_d(t,\mathbf{S}) &:= \sum_{i=1}^{n} P_d(t,S_i)\tau_K(S_{i-1},S_i),
\end{aligned} \tag{31}$$

where $\omega = -1/1, R_x^{\text{Swap}}(t;\mathbf{T},\mathbf{S}), A_d(t,\mathbf{S}), F_{x,j}(t)$ are the swap side (receiver/payer of the fixed rate), the swap rate, the annuity, and the FRA rate, respectively, of the Swap.

**Forward Rate Agreements** (FRA) are elementary Swaps with a single cash flow on both legs.

Interest rate **Basis Swaps** (Figure 7) are floating vs floating Swaps tied to two Libors with different tenors. The Basis Swap schedule and price in terms of basis swap spread are given, using eq. (31), by

$$\begin{aligned}
&\mathbf{S} = \{S_0,...,S_n\}, \text{ fixed leg schedule,} \\
&\mathbf{T}_x = \{T_{x,0},...,T_{x,m_x}\}, \text{ floating leg } x \text{ schedule,} \\
&\mathbf{T}_y = \{T_{y,0},...,T_{y,m_y}\}, \text{ floating leg } y \text{ schedule,} \\
&S_0 = T_{x,0} = T_{y,0},\ S_n = T_{x,m_x} = T_{y,m_y}, \\
&\Delta(t;\mathbf{T}_x,\mathbf{T}_y,\mathbf{S},\omega) := R_x^{\text{Swap}}(t;\mathbf{T}_x,\mathbf{S}) - R_y^{\text{Swap}}(t;\mathbf{T}_y,\mathbf{S}) \\
&= \frac{\sum_{i=1}^{m_x} P_d(t;T_{x,i})F_{x,i}(t)\tau_x(T_{x,i-1},T_{x,i}) - \sum_{j=1}^{m_y} P_d(t;T_{y,j})F_{y,j}(t)\tau_y(T_{y,j-1},T_{y,j})}{A_d(t;\mathbf{S})}.
\end{aligned} \tag{32}$$

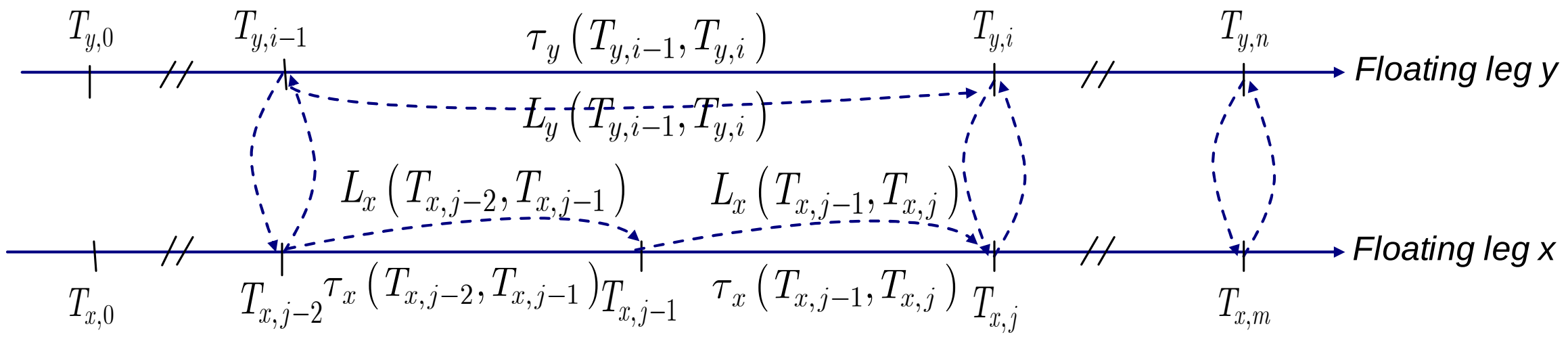

**Figure 7:** schematic picture of a Basis Swap contract, with tenor y=2x, e.g. Libor 6M vs Libor 3M.

The basis spread is defined as the difference between the swap rates of two fixed vs floating Swaps with equal fixed legs (annual payment frequency) and with floating legs indexed to the two Libors with different tenors (and corresponding payment frequencies). FRA and Swap contracts for different currencies, tenors and maturities are quoted in real time on the financial market. Using the formulas above, written in terms of the FRA rates $F_{x,j}(t)$, the corresponding FRA rate term structures can be extracted from such quotations and used, by interpolation, to price similar FRAs/Swaps not directly quoted on the market.

In the limit of negligible counterparty/liquidity risk, eq. (28), we return to a single-curve framework (the indexes *d*, *x* drops) and we obtain the classical Swap pricing expression cited in pre-crisis textbooks on interest rate modelling (e.g. [21], [22]):

$$\begin{aligned}
\mathbf{Swap}_{\text{float}}(t;\mathbf{T}) &= N\sum_{j=1}^{m} P_d(t,T_j)F_{x,j}(t)\tau_x(T_{j-1},T_j) \\
&\to N\sum_{j=1}^{m} P(t,T_j)F_j(t)\tau(T_{j-1},T_j) \\
&= N\sum_{j=1}^{m} P(t,T_j)\left[\frac{P(t,T_{j-1})}{P(t,T_j)}-1\right] \\
&= N\left[P(t,T_0)-P(t,T_m)\right].
\end{aligned} \tag{33}$$

## References

[1] See Zeeman's biography at the Nobel Prize website *http://nobelprize.org* and Wikipedia[5] at *http://en.wikipedia.org/wiki/Pieter_Zeeman*.

[2] See P. Zeeman, "*The Effect of Magnetisation on the Nature of Light Emitted by a Substance*", Nature 55, p. 347, 11 February 1897; R. P. Feynman, R. Leighton, M. Sands, "*The Feynman Lectures on Physics*", vol. 3, ch. 12-4; Wikipedia at *http://en.wikipedia.org/wiki/Zeeman_effect*.

[3] D. Wood, "*Libor fix*", Risk Magazine, 1 Jul. 2011.

[4] C. Snider, T. Youle, "*Does the Libor reflect banks' borrowing costs ?*", 2 Apr. 2010, SSRN working paper, *http://ssrn.com/abstract=1569603*

[5] M. Morini, "*Solving the Puzzle in the Interest Rate Market*", Oct. 2009, SSRN working paper, *http://ssrn.com/abstract=1506046*.

[6] B. Tuckman and P. Porfirio, "*Interest Rate Parity, Money Market Basis Swap, and Cross-Currency Basis Swap*", Lehman Brothers Fixed Income Liquid Markets Research – LMR Quarterly , 2004, Q2.

[7] ISDA, "*ISDA Margin survey 2011*", 14 Apr. 2011, *http://www.isda.org*.

[8] V. Piterbarg, "*Funding beyond discounting: collateral agreements and derivatives pricing*", Risk, Feb. 2010.

[9] M. Morini, A. Prampolini, "*Risky Funding with counterparty and liquidity charges*", Risk, Mar. 2011.

[10] C. Burgard, M. Kjaer, "*In the Balance*", 14 Mar. 2011, SSRN working paper, http://ssrn.com/abstract=1785262

[11] C. Fries, "*Discounting Revisited – Valuations under Funding Costs, Counterparty Risk and Collateralization*", 15 May 2010, SSRN working paper, *http://ssrn.com/abstract=1609587*.

[12] A. Castagna, ""*Funding, Liquidity, Credit and Counterparty Risk: Links and Implications*", DefaultRisk.com working paper, *http://www.defaultrisk.com/pp_liqty_53.htm*.

[13] M. Fujii and A. Takahashi, "*Choice of Collateral Currency*", Risk, Jan. 2011.

[14] F. Ametrano, M. Bianchetti, "*Bootstrapping the Illiquidity: Multiple Yield Curves Construction For Market Coherent Forward Rates Estimation*", in "*Modeling Interest Rates: Latest Advances for Derivatives Pricing*", edited by F. Mercurio, Risk Books, 2009.

[15] H. Lipman, F. Mercurio, "*The New Swap Math*", Bloomberg Markets, Feb. 2010.

[16] M. Bianchetti, "*Two Curves, One Price*", Risk, August 2010.

[17] M. Bianchetti, M. Carlicchi "*Interest Rates after the Credit Crunch: Multiple Curve Vanilla Derivatives and SABR*", SSRN working paper, *http://ssrn.com/abstract=1783070*.

[18] N. Sawyer, "*ISDA working group to draw up new, standardised CSA*", Risk, 15 Feb. 2011.

[5] We remind that Wikipedia is a remarkable secondary (not primary) source of information.

[19] D. Brigo, A. Capponi, A. Pallavicini, V. Papatheodorou, "*Collateral Margining in Arbitrage-Free Counterparty Valuation Adjustment including Re-Hypotecation and Netting*", SSRN working paper, *http://ssrn.com/abstract=1744101*.
[20] F. Mercurio, "*Modern LIBOR Market Models: Using Different Curves for Projecting Rates and for Discounting*", International Journal of Theoretical and Applied Finance, Vol. 13, No. 1, 2010, pp. 113-137.
[21] D. Brigo and F. Mercurio, "*Interest Rate Models: Theory and Practice*", 2nd edition, 2006, Springer.
[22] Leif B.G. Andersen and Vladimir V. Piterbarg, "*Interest Rate Modeling*", 1st edition, 2010, Atlantic Financial Press.